\documentclass[aps, preprint, tightenlines, showpacs,nofootinbib,superscriptaddress, 12pt]{revtex4}

\usepackage{amssymb}
\usepackage{amsmath}
\usepackage{epsfig}

\begin{document}

\title{Probing Transverse Momentum Broadening in Heavy Ion Collisions}

\author{A. H. Mueller}
\affiliation{Department of Physics, Columbia University, New York, NY 10027, USA}

\author{Bin Wu}
\affiliation{Department of Physics, The Ohio State University, Columbus , OH 43210, USA}
\affiliation{Institut de Physique Theorique, CEA Saclay, UMR 3681, F-91191 Gif-sur-Yvette, France}

\author{Bo-Wen Xiao}
\affiliation{Key Laboratory of Quark and Lepton Physics (MOE) and Institute
of Particle Physics, Central China Normal University, Wuhan 430079, China}

\author{Feng Yuan}
\affiliation{Nuclear Science Division, Lawrence Berkeley National
Laboratory, Berkeley, CA 94720, USA}

\begin{abstract}
We study the dijet azimuthal de-correlation in relativistic heavy ion collisions as an important
probe of the transverse momentum broadening effects of a high 
energy jet traversing the quark-gluon plasma. 
We take into account both the soft gluon radiation in vacuum associated with the Sudakov 
logarithms and the jet $P_T$-broadening effects in the QCD
medium. We find that the Sudakov effects are dominant at the LHC, while
the medium effects can play an important role at RHIC energies. This explains why the LHC
experiments have not yet observed sizable $P_T$-broadening effects in the measurement of dijet azimuthal correlations
in heavy ion collisions. Future investigations at RHIC will provide a unique opportunity to 
study the $P_T$-broadening effects and help to pin down the 
underlying mechanism for jet energy loss in a hot and dense medium.
\end{abstract}
\maketitle

{\it Introdcution.}
One of the most important discoveries in the relativistic heavy ion 
experiments at RHIC at Brookhaven National Laboratory and the
LHC at CERN is the jet quenching phenomena~\cite{Adcox:2001jp,Adler:2002xw,Aad:2010bu,Chatrchyan:2011sx}, 
where high energy partons lose tremendous energy through their interactions with the quark-gluon
plasma created in heavy ion collisions. Theoretically, the jet energy loss can be understood as a result of the induced gluon radiation when the 
parton traverses the hot QCD matter, and has been well formulated in the QCD
framework~\cite{Gyulassy:1993hr,Baier:1996kr,Baier:1996sk, Baier:1998kq, Zakharov:1996fv}. Alternatively, 
the strong coupling feature of the medium can be described by models based on the Ads/CFT
correspondence in string theory~\cite{Herzog:2006gh, Gubser:2006bz, Liu:2006ug, CasalderreySolana:2006rq, CasalderreySolana:2011us,Dominguez:2008vd}. These calculations have been successfully applied to heavy ion phenomenology in order to understand the jet quenching related 
experimental data from the RHIC and LHC~\cite{Burke:2013yra}.

Meanwhile, there has been a strong theoretical argument that the jet energy loss is 
associated with the $P_T$-broadening phenomena~\cite{Baier:1996sk}, where 
the energetic jet accumulates additional transverse momentum perpendicular 
to the jet direction.  Combining the analysis of the jet energy loss and 
$P_T$-broadening is of crucial importance to 
consolidate the underly mechanism for the jet energy loss.
Dijet production is an idea process for this physics, where we can use the leading
jet as a reference. The jet energy loss can be studied by measuring the energy of the away side
jet, and the $P_T$-broadening effects can be accessed through the 
azimuthal angular correlation. The former has been investigated by the ATLAS
and CMS collaborations at the LHC through the so-called $A_J$ distribution
measurements, where the theoretical interpretations are consistent with the jet 
energy loss~\cite{Qin:2010mn,Young:2011qx,He:2011pd}. Similar conclusion has been 
reached also for photon-jet events~\cite{Chatrchyan:2012gt}, see, e.g., Ref.~\cite{Wang:2013cia}.
Both experiments have also studied the azimuthal angular correlation between the two jets, but found no difference
as compared to the $pp$ collisions. The goal of this paper is to perform
a systematic study on the dijet azimuthal de-correlation in heavy ion collisions. In 
particular, we find that the $P_T$-broadening effects plays a negligible role at the 
LHC energy, whereas, it will become an important contribution and
should be observed at the RHIC energy, since Sudakov effects at the LHC are much stronger than that at RHIC.
The experimental investigation of this $P_T$-broadening effects in dijet 
production is a crucial step forward to identify the underlying mechanism for the 
jet energy loss in heavy ion collisions.

\begin{figure}[tbp]
\begin{center}
\includegraphics[height=6cm]{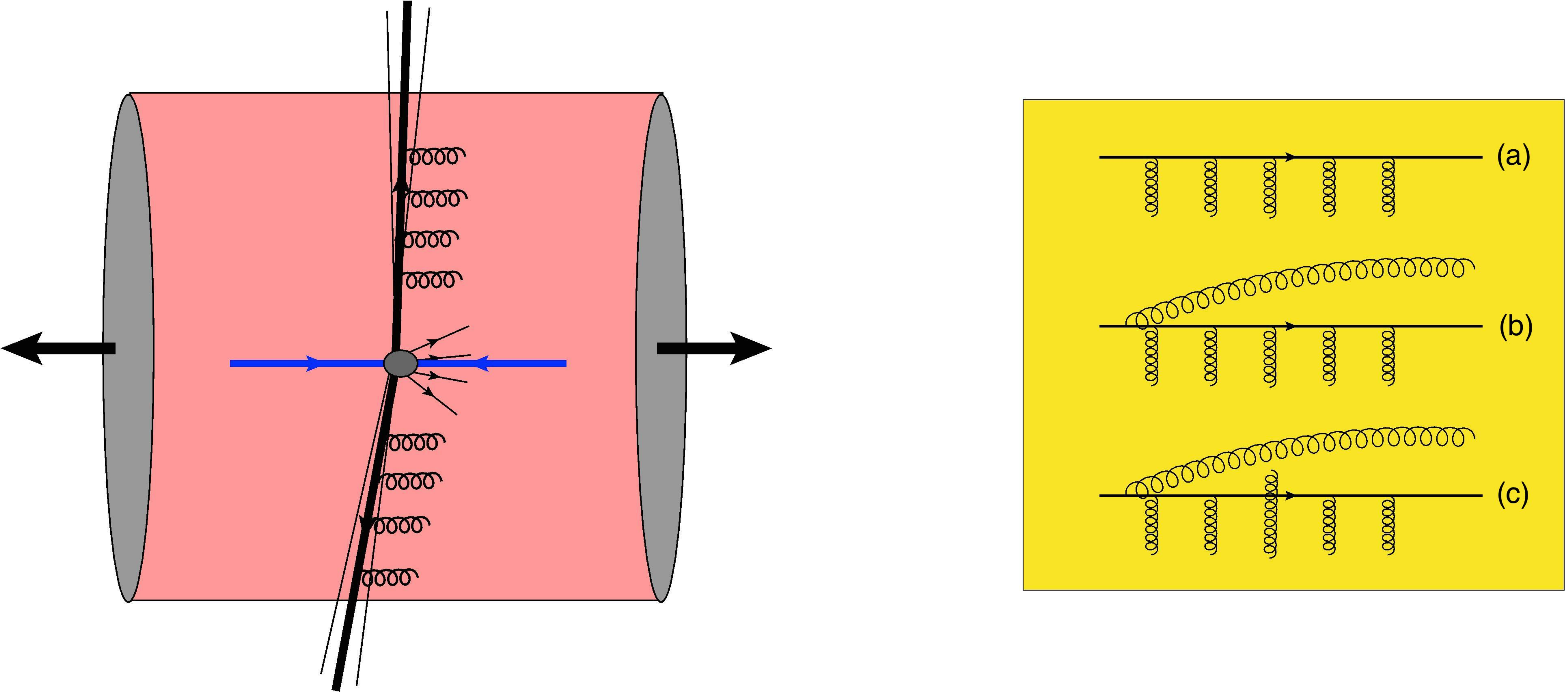}
\end{center}
\caption[*]{Dijet production and azimuthal angular de-correlation in heavy ion
collisions: both soft gluon radiation when the two jets are produced from the 
partonic scattering processes and the multiple scattering between the high energy
jet and the medium induced gluon radiation contribute to the de-correlation.}
\label{dijet}
\end{figure}

As illustrated in Fig.~\ref{dijet}, dijets are produced in partonic scattering, which 
go through the hot QCD medium before reaching the detector,
\begin{equation}
A+A\to Jet_1+Jet_2+X\ .
\end{equation}
Most of the dijet events are produced in the back-to-back azimuthal correlation configuration
with the azimuthal angle: $\Delta \phi=\phi_1-\phi_2\sim \pi$,
where $\phi_{1,2}$ are the azimuthal angles of these two final state jets with 
transverse momenta $k_{1\perp}$ and $k_{2\perp}$, respectively. 
There are two important contributions to the azimuthal de-correlation 
of the two jets in heavy ion collisions:  one is the soft and collinear gluon 
radiation associated with the partonic $2\to 2$ subprocesses, which is 
referred to as the Sudakov effects;  the other is the $P_T$-broadening effects due to multiple scattering and medium 
induced radiation when high energy jets propagate through the medium. 
Therefore, in order to unequivocally determine the $P_T$-broadening
effects from the medium, we have to first understand the Sudakov effects in dijet
production. Because this comes from the partonic scattering, we can study it
in dijet production in $pp$ collisions, which has been extensively investigated 
by the experiments at the Tevatron\cite{Abazov:2004hm} and the LHC~\cite{Khachatryan:2011zj,daCosta:2011ni}. 
The theoretical developments\cite{Mueller:2013wwa,Sun:2014gfa} in the last few years have also advanced, where a successful 
description of these data was found~\cite{Sun:2014gfa}.
In the following, we will compare the relative importance of the Sudakov and 
$P_T$-broadening effects at the LHC and RHIC. 
This will provide a benchmark calculation for the $P_T$-broadening
in high energy hard scattering processes in $AA$ and $pA$ collisions.

{\it Sudakov and $P_T$-broadening effects.}
We follow the BDMPS framework\cite{Baier:1996kr, Baier:1996sk, Baier:1998kq} to analyze the $P_T$-broadening effects in 
the heavy ion collisions, and compare that with the Sudakov effects from 
gluon radiation in vacuum. As illustrated in the right panel of Fig.~\ref{dijet} (a),
when a high energy jet traverses the medium, it suffers multiple
scatterings and medium induced gluon radiation. These effects can be represented by a characteristic scale $Q_s^2=\hat qL$, 
which depends on the transport coefficient $\hat q$~\cite{Baier:1996sk}, and the length of the jet path in the medium $L$.
The physics behind the $P_T$-broadening is 
that each scattering randomly gives a small transverse momentum kick to the jet,
which in turn accumulates a total transverse momentum of order $Q_s$
along the path in the medium. In the BDMPS framework, 
this effects is computed in a Glauber multiple scattering
theory, and the result is expressed in the Fourier transformation 
conjugate $b_\perp$-space as $e^{-{Q_s^2b_\perp^2}/{4}}$. 
When Fourier transforming back to the transverse momentum
space, it leads to a Gaussian-like distribution of $e^{-q_\perp^2/Q_s^2}$,
where $q_\perp$ represents the transverse momentum perpendicular 
to the jet direction. In addition, 
recent studies\cite{Wu:2011kc, Liou:2013qya, Iancu:2014kga, Blaizot:2014bha, Wu:2014nca} 
reveal that additional medium induced gluon radiation can also contribute to the 
jet $P_T$-broadening, and leads to slightly larger values of $Q_s^2$. 

The numerical $\hat q$ parameter has been a subject of intensive studies
in jet quenching phenomenology, see a recent report from  
the JET-collaboration~\cite{Burke:2013yra}, which gives roughly 
$Q_s^2=\hat q L\simeq6 \,\textrm{GeV}^2$ at RHIC and 
$\hat q L\simeq 10\, \textrm{GeV}^2$ at the LHC for quark 
jets and medium length $L=5\, \textrm{fm}$. For gluon jets, 
$\tilde{Q}_s^2$ is $\frac{2N_c^2}{N_c^2 -1}$ times of that 
for quark jets due to different Casimir factors. 

The medium related $P_T$-broadening effect is physically different from the 
Sudakov effects computed from the collinear and soft
gluon radiation in hard scattering processes~\cite{Collins:1984kg}. 
To see this more clearly, we compare the effects
from the gluon radiation contributions in the right panel of Fig.~\ref{dijet}.
The vacuum radiation diagram of (b) has been excluded in the medium
induced radiation contribution in the BDMPS calculations, which, 
on the other hand, is part of the collinear and soft gluon radiation contribution 
to the imbalance between the two jets in the dijet production
process. In particular, this final state gluon radiation
will contribute to a term depending the jet size~\cite{Sun:2014gfa}:
$\frac{\alpha_s}{2\pi^2}\frac{1}{q_\perp^2}C_f\ln\frac{1}{R^2}\ ,$
where $q_\perp$ represents the transverse momentum of the radiated
gluon, $R$ the jet size, and $C_f$ the color factor for
the associated jet ($C_F$ for quark jet and $C_A$ for gluon jet). 
This contribution can be factorized into the soft factor
in the dijet production. When we Fourier transform
the above expression into $b_\perp$-space, we obtain a 
logarithmic dependence $\ln (b_\perp^2\mu^2)$. In the factorization
formula, the scale $\mu$ will be set around the
hard momentum scale (such as the leading jet energy) to resum 
the associated large logarithms. 

On the other hand, when we consider the medium induced gluon radiation 
from the diagram (c) of the right panel of Fig.~\ref{dijet}, it is an infrared
safe contribution as demonstrated in the BDMPS calculation~\cite{Baier:1996sk}. 
This is because the famous Landau-Pomeranchuk-Migdal (LPM) effect suppress the small 
transverse momentum gluon radiation in the medium. 
There is no such $1/q_\perp^2$ behavior from this diagram, and it does 
not contribute to a logarithmic term of $\ln(b_\perp^2\mu^2)$,
although it will contribute to a high order corrections to $Q_s$~\cite{Wu:2011kc,Liou:2013qya,tobe}.
The bottom line is that the Sudakov effects only take into account the gluon radiation in the vacuum which contains 
no information of medium effects. In the meantime, it is common practice to subtract the vacuum contribution when 
we compute any medium effects in the BDMPS formalism. That also indicates that we can 
factorize these two effects into Sudakov factor and medium-dependent quantities, respectively, and write them together in a unified 
formalism to describe the dijet azimuthal correlation. We plan to discuss the separation of these two effects in detail in a separate publication~\cite{tobe}.

The vacuum radiation diagram is part of all collinear and soft
gluon radiation contributions in dijet production, which has
been calculated recently in Refs.~\cite{Mueller:2013wwa,Sun:2014gfa}.
From these calculations, it was found that there is a simple
power counting rule, which allows us to predict that each incoming parton contributes
to the leading double logarithm with the associated color factor. 
They can be factorized into the so-called transverse momentum 
distributions from the incoming nucleons and the soft factor associated
with final state jets. Because of the short distance hard scattering for dijet 
production, these contributions will not be modified in heavy ion collisions. 
We extend the resummation formula derived in Ref.~\cite{Sun:2014gfa} in our study as follows
\begin{eqnarray}
&&\frac{d^4\sigma}{dy_1 dy_2 d k_{1\perp}^2
d^2k_{2\perp}}=\sum_{ab}\sigma_0\int\frac{d^2\vec{b}_\perp}{(2\pi)^2}
e^{-i\vec{q}_\perp\cdot \vec{b}_\perp}W(b_\perp)\ , \label{xc}
\end{eqnarray}
where we focus in the small $q_\perp\ll k_{1\perp}\sim k_{2\perp}$ region. 
Away from this region, we have to include a fixed order perturbative correction. In the low
$q_\perp$ region, we apply an all order resummation formula for 
$W(b_\perp)$,
\begin{eqnarray}
&&W(b_\perp)=x_1\,f_a(x_1,\mu_b)
x_2\, f_b(x_2,\mu_b) e^{-S(Q^2,b_\perp)}\  ,\label{wb}
\end{eqnarray}
where $\sigma_0$ represents normalization of the
differential cross section, $y_1$ and $y_2$ are rapidities of the two jets,
$Q^2=\hat s=x_1x_2S$ is the partonic center of mass energy squared, $b_0=2e^{-\gamma_E}$,
$f_{a,b}(x,\mu_b=b_0/b_*)$ are parton distributions for the incoming partons $a$ and $b$, 
$x_{1,2}=k_{1\perp}\left(e^{\pm y_1}+e^{\pm y_2}\right)/\sqrt{S}$
are momentum fractions of the incoming hadrons carried by the partons.
By introducing the $b_*$-prescription\cite{Collins:1984kg} which sets $b_*=b_\perp /\sqrt{1+b^2_\perp/b_{max}^2}$ with $b_{max}=0.5 \textrm{GeV}^{-1}$, we separate the Sudakov form factor $S(Q,b_\perp)$ 
into perturbative and non-perturbative parts in $pp$ collisions: 
$S(Q,b_\perp)=S_{pert}(Q,b_\perp)+S_{NP}(Q,b_\perp)$ with the perturbative part defined as,
\begin{eqnarray}
S_{pert}(Q^2,b_\perp)&=&\int^{Q^2}_{\mu_b^2}\frac{d\mu^2}{\mu^2}
\left[A\ln\left(\frac{Q^2}{\mu^2}\right)+B+(D_1+D_2)\ln\frac{1}{R^2}\right]\ , \label{su}
\end{eqnarray}
where $R$ represents the jet size. We have applied the anti-$k_t$
algorithm to define the final state jets in our calculations. 
Here the coefficients $A$, $B$, $D_1$, $D_2$ can be expanded
perturbatively in terms of powers of $\alpha_s$. At one-loop order, 
$A=C_A \frac{\alpha_s}{\pi}$,
$B=-2C_A\beta_0\frac{\alpha_s}{\pi}$ for gluon-gluon initial state,
$A=C_F \frac{\alpha_s}{\pi}$,
$B=\frac{-3C_F}{2}\frac{\alpha_s}{\pi}$ for quark-quark initial state,
and $A=\frac{(C_F+C_A) }{2}\frac{\alpha_s}{\pi}$,
$B=(\frac{-3C_F}{4}-C_A\beta_0)\frac{\alpha_s}{\pi}$ for gluon-quark initial state.
$D_i$ is $\frac{\alpha_s}{2\pi} C_F$ for quark jet, and $\frac{\alpha_s}{2\pi}$ for gluon jet. 
For the non-perturbative part, we follow those in Ref.~\cite{Sun:2014gfa}.

{\it Probing the $P_T$-broadening effect.} 
In the following dicussion, we focus on dijet productions in mid-rapidity. In this kinematics,
the $P_T$-broadening effects also contribute to the longitudinal momentum 
along the incoming beam direction, and therefore modify the rapidity of the final state jets. 
However, in the region of interest for our study, this is a sub-leading
order effects, which can be neglected in our calculations. 
We also notice that the $P_T$-broadening effect is along the direction
perpendicular to the jet, so that it will not affect the transverse momentum
along the jet direction. However, for convenience in implementing
the $P_T$-broadening effects in a single formula together with the Sudakov
resummation, we modify Eq.~(\ref{su}) as 
\begin{equation}
S(Q,b)|_{AA}=S_{pert}(Q,b_*)+S_{NP}(Q,b)+{Q}_s^2b^2/4 \ , \label{formfactor}
\end{equation}
where $Q_s^2$ encodes the medium $P_T$ broadening effects. 
In the correlation calculation, we will integrate out the sub-leading jet energy in a certain range,
which effectively integrates out the transverse momentum along the jet direction.
As a result, Eq.~(\ref{formfactor}) is reduced to the form that the $P_T$ 
broadening effects only apply to the transverse direction perpendicular to the jet direction. 

The first two terms in Eq.~(\ref{formfactor}) are the same as that in $pp$ collisions~\footnote{
Here we neglect the $P_T$-broadening from the cold nuclei effects, which
is much smaller than that in hot QCD matter~\cite{Baier:1996sk,Burke:2013yra}.}. 
In order to observe the $P_T$-broadening effects, we have to find the right 
kinematics where the last term will be important. This can be achieved 
by varying the jet energy (which will modify the perturbative 
Sudakov term) or the medium effects (by changing the centrality or the
energy of the collisions).

\begin{figure}[tbp]
\begin{center}
\includegraphics[height=6.0cm]{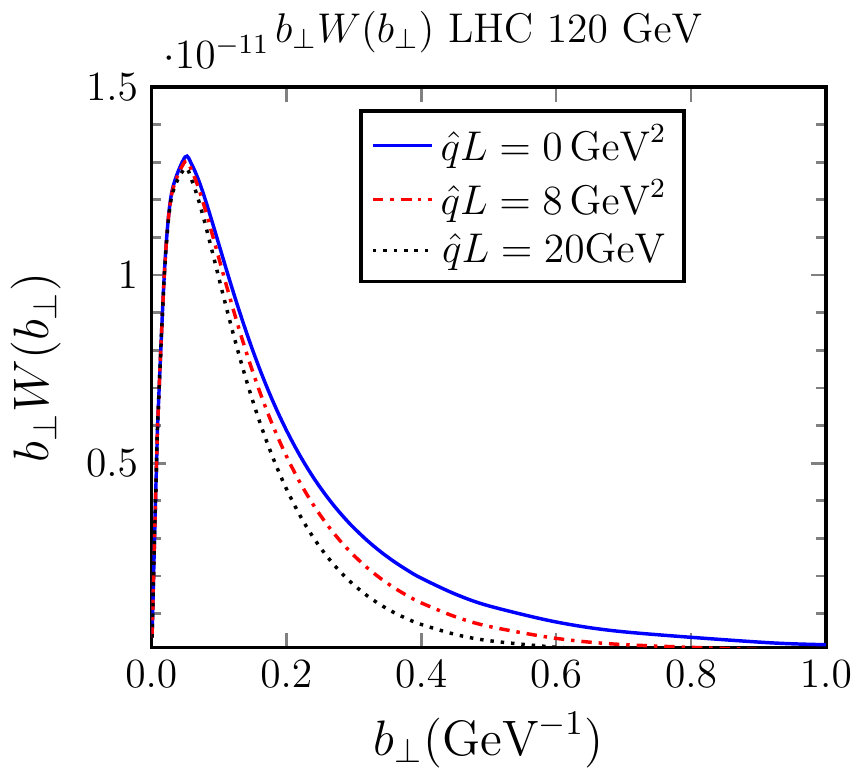}
\includegraphics[height=6.0cm]{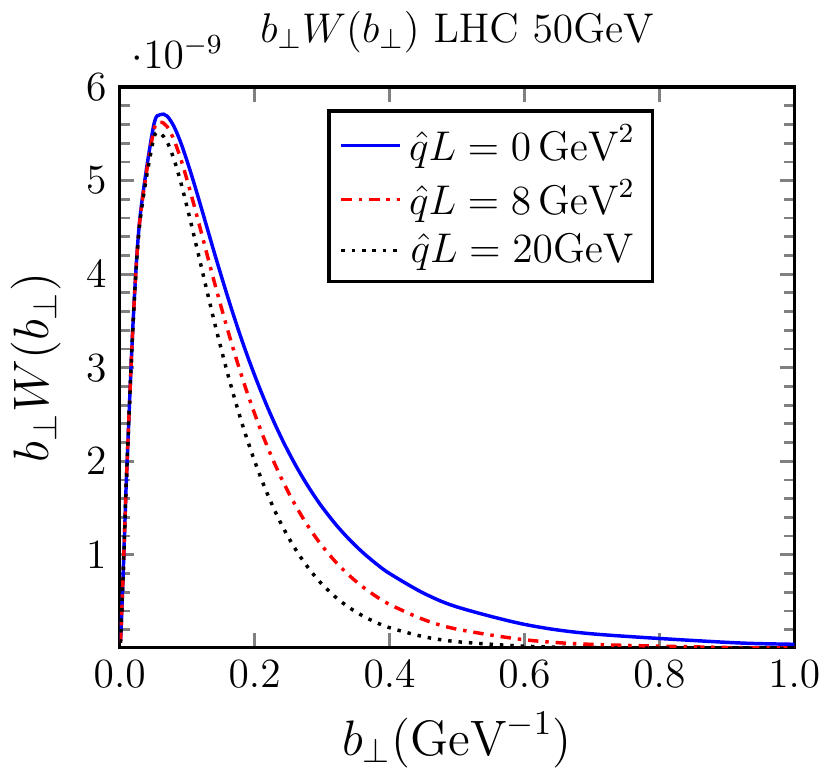}
\end{center}
\caption[*]{Impact of the $P_T$-broadening effects on dijet production at mid-rapidity
at the LHC, where we plot the $b_\perp \times W(b_\perp)$ of Eq.~(\ref{wb}) as functions 
of $b_\perp$ with $S(Q,b)$ in Eq.~(\ref{formfactor}) and three different
values of ${Q}_s^2=0,8,20$GeV${}^2$. 
The Fourier transformation of $W(b_\perp)$ would give the imbalance 
transverse momentum $\vec{q}_\perp=\vec{k}_{1\perp}+\vec{k}_{2\perp}$ distributions,
where $k_{1\perp}$ and $k_{2\perp}$ are the leading jet and sub-leading
jet transverse momenta. Comparison between the two choices of the leading 
jet transverse momentum $P_\perp=120$,$50$GeV at the LHC, 
respectively.}
\label{di-cms}
\end{figure}

\begin{figure}[tbp]
\begin{center}
\includegraphics[height=6cm]{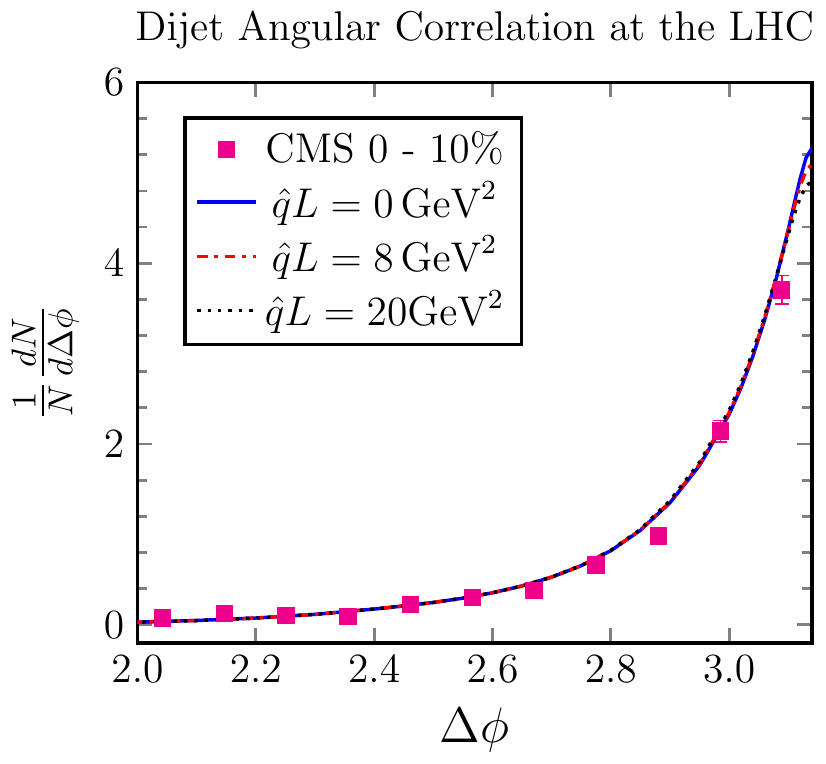}
\end{center}
\caption[*]{$P_T$-broadening effects in Dijet azimuthal angular
distributions in central PbPb collisions at the LHC. 
}
\label{dijetp}
\end{figure}

Let us first examine the typical dijet production at the LHC. In 
Fig.~\ref{di-cms}, we plot $b_\perp \times W(b_\perp)$ as function of $b_\perp$ for a leading
jet energy $P_\perp=120$ and $50$GeV, respectively, at mid-rapidity at $\sqrt{S}=2.76TeV$,
where $W(b_\perp)$ is defined as in Eq.~(\ref{wb}). The Fourier transformation of
$W(b)$ yields the imbalance $q_\perp$ distribution for the dijet.
In the numeric calculations, we have taken into account the perturbative
form factor at one-loop order: $A^{(1)}$, $B^{(1)}$, and $D^{(1)}$. 
We have also checked the complete next-to-leading logarithmic corrections
do not change significantly the behavior of these distributions.
The three curves in this plot correspond to ${Q}_s^2=0,8,20$GeV${}^2$, respectively.
From these plots, we can see that the dominant contribution
of $W(b_\perp)$ comes from small-$b$ region, where the $P_T$-broadening 
effects do not affect the results at all. 
Clearly, at the LHC, the perturbative Sudakov form factor $S_{pert}(b)$
dominates the small-$b$ contribution. More importantly, in the LHC energy region, 
the dijet productions probe relatively small-$x$ parton distributions, where
the $xf_a(x,\mu_b)$ factor in Eq.~(\ref{wb}) significantly push the contributions
into the small-$b$ region. Therefore, even if we lower the leading jet 
energy to $50\textrm{GeV}$, it will still be dominated by small-$b$ contribution as shown in
the right panel of Fig.~\ref{di-cms}, where, again, we find that
the medium effects are negligible.

To see the medium effects on the azimuthal angular distribution, we apply 
Eqs.~(\ref{xc}, \ref{wb}, \ref{su}, \ref{formfactor}) to calculate the $\Delta \phi$ distribution,
\begin{equation}
\frac{1}{\sigma_{dijet}}\frac{d\sigma_{dijet}}{d\Delta\phi} \ ,
\end{equation}
where $\sigma_{dijet}$ is the dijet cross-section and
the numerator is calculated from Eq.~(\ref{xc}) after integrating over other kinematic
variables. As shown in Fig.~\ref{dijetp}, we find that the shape of the angular correlation 
is consistent with the CMS data for back-to-back dijet configurations. 
More importantly, our results show that
$P_T$-broadening effects are negligible at the LHC, where 
the three curves (corresponding to three different choices
for $Q_s$) almost lay on top of each other. 
This also explains why the azimuthal angular correlation in
dijet productions does not change from $pp$ to $AA$ collisions
at the LHC for the kinematical region studied in the ATLAS and CMS
measurements.

\begin{figure}[tbp]
\begin{center}
\includegraphics[height=6.0cm]{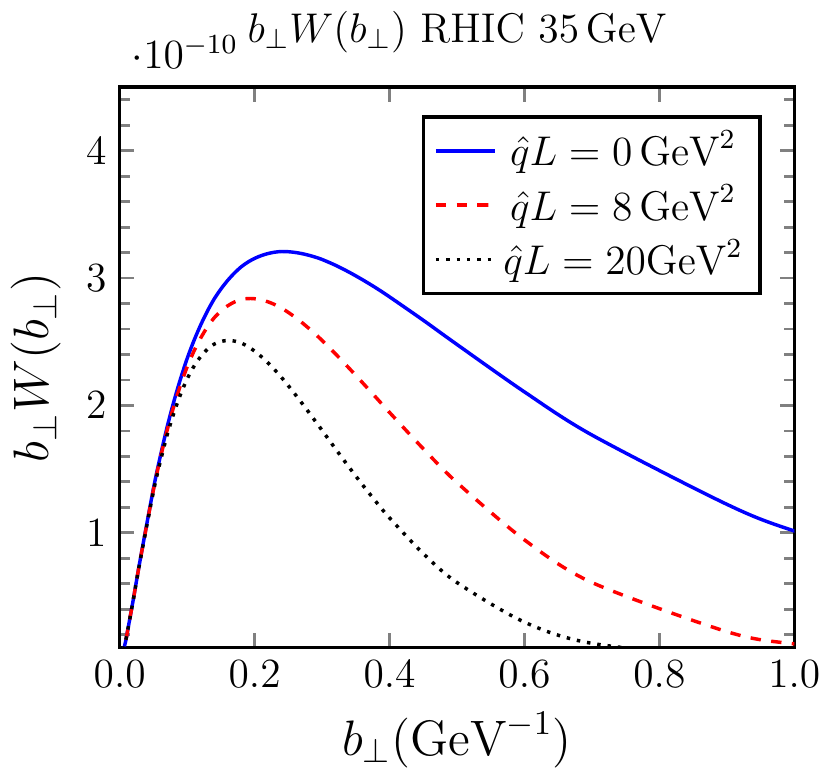}
\includegraphics[height=6.0cm]{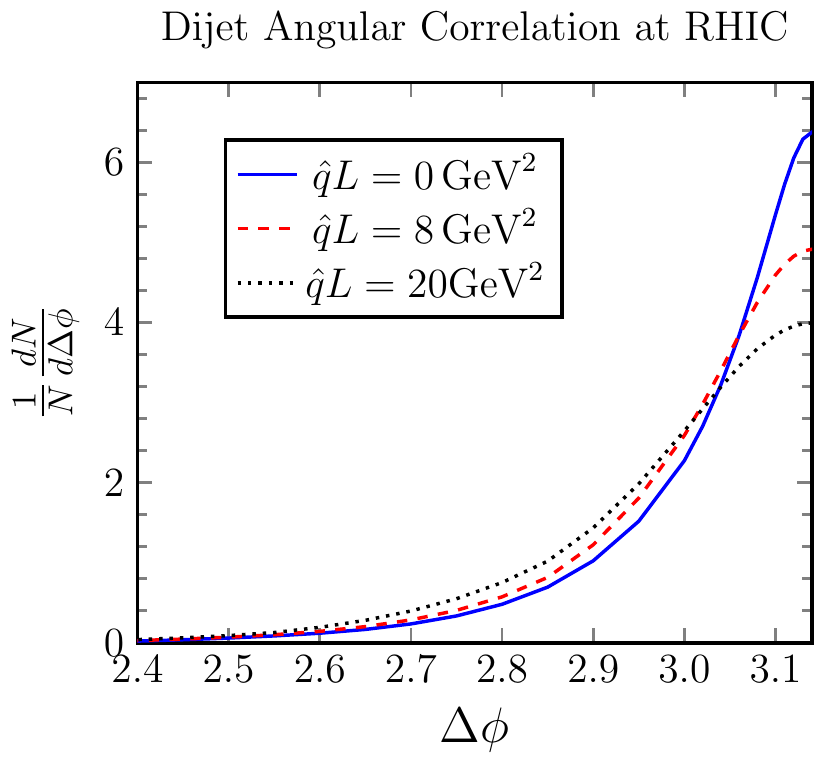}
\end{center}
\caption[*]{$P_T$-broadening effects at RHIC: (left) plot of $b_\perp W(b_\perp)$ as function of
$b_\perp$; (right) azimuthal de-correlation for dijet production
at RHIC for a leading jet $P_\perp=35$GeV.}
\label{dijetrhic}
\end{figure}

Nevertheless, the above conclusions can dramatically 
change when we switch from the LHC to RHIC. As shown in Fig.~\ref{dijetrhic}, we plot the same distributions
for a typical dijet production at RHIC with $\sqrt{S}=200$GeV. Here, clearly,
we can see that the medium induced $P_T$-broadening contribution is very important
in the $b\sim 0.5$GeV${}^{-1}$ region. 
As a result, significant $P_T$-broadening effects can be found
in Fig.~\ref{dijetrhic} for RHIC experiments. 
In particular, the $P_T$
broadening effects changes not only the shape but also the magnitude
of the dijet azimuthal correlations in heavy ion collisions at RHIC.
We are looking forward to these measurements in the near 
future~\cite{Adare:2015kwa}.

{\it Conclusions.}
We have performed a systematic study of dijet azimuthal de-correlation 
in heavy ion collision to probe the $P_T$-broadening effects in the quark-gluon
plasma. By taking into account additional Sudakov effects, we found
that at the LHC, the medium $P_T$-broadening effects are negligible in the dijet
azimuthal angular distribution, which is consistent with the observations
from the ATLAS and CMS experiments. By contrast, we demonstrated
that the $P_T$-broadening effects can be important at the RHIC energy
and we should be able to observe it in experiments.
Future study of this physics at RHIC would provide a unique
opportunity to probe the $P_T$-broadening effects and help 
to identify the underlying mechanism for the jet energy loss in
relativistic heavy ion collisions.

We also note that there have been significant progress in the development of 
the Monte Carlo event generator ``JEWEL"\cite{Zapp:2008gi, Zapp:2012ak}, 
which incorporates both the parton shower effects 
and medium effects, such as the LPM effects. By and large, our theoretical work is complementary to these 
numerical studies.

Further theoretical investigations should follow along the direction of this paper.
In a recent calculation at the next-to-leading order~\cite{Liou:2013qya},
a double logarithmic term depending on the length $L$ was found in 
$Q_s$. Since we are dealing with jet 
propagation in this paper, we need to consider the modification of the finite jet size on the
$P_T$-broadening calculations. These contributions will depend
on the details of gluon radiation in the medium and could provide an unique
way to distinguish different mechanisms. We should also combine the 
above analysis with the jet energy loss calculations.
For that, we need to carry out a next-to-leading order perturbative calculation
combined with the jet energy loss with jet size dependence (see, e.g, recent
calculations in Ref.~\cite{Wu:2014nca, Ghiglieri:2015ala}). 
Early attempts have been made in Refs.~\cite{Qin:2010mn,Young:2011qx,He:2011pd} 
to calculate the $A_J$ asymmetries from the LHC measurements. 
By combining the theoretical studies of the azimuthal de-correlation and the energy
asymmetry $A_J$ for dijet production in heavy ion collisions, together with the sophisticated Monte Carlo simulations, 
we should be able to unambiguously decode the underlying mechanism for 
jet quenching phenomena in the strongly interaction
quark-gluon plasma.

We thank Y.J. Lee, G.Y. Qin, J.W. Qiu, X.N. Wang for interesting
discussions and comments. We also thank A. Angerami and B. Cole for useful and interesting discussions
on the subject matter of this paper. This work was supported in part by the U.S. Department of 
Energy under the contracts DE-AC02-05CH11231 and by the NSFC under Grant No.~11575070.

\end{document}